\documentclass[english, runningheads, pdfusetitle]{llncs}[2018/03/10]

\newif\ifLNCSAnonymous

\usepackage{stefan2020}

\usepackage[activate={true,nocompatibility},final,tracking=true,kerning=true,spacing=true,factor=1100,stretch=00,shrink=25]{microtype}

\begin{document}

\title{Trust Me If You Can: Trusted Transformation Between (JSON) Schemas to Support Global Authentication of Education Credentials\thanks{This work was supported by the European Union's Horizon 2020 Framework Programme for Research and Innovation under grant agreement \textnumero~871473 (KRAKEN) as well as the Josef Ressel Center for Blockchain Technologies and Security Management (JRC Blockchains).
The authors would also like to thank TU Graz' Registrar's Office for insights into verification of (paper-based) diplomas.
}}

\titlerunning{Trusted Transformation Between (JSON) Schemas}


\author{Stefan More\inst{1}\orcidID{0000-0001-7076-7563} \and
Peter Grassberger\inst{1,2} \and\\
Felix Hörandner\inst{1}\orcidID{0000-0001-8591-3463} \and
Andreas Abraham\inst{1}\orcidID{0000-0002-4163-9113} \and
Lukas Daniel Klausner\inst{3}\orcidID{0000-0003-3650-9733}}

\authorrunning{More, Grassberger, Hörandner, Abraham, Klausner}

\institute{Graz University of Technology, Graz, Austria \\
\email{\{smore,fhoerandner,aabraham\}@iaik.tugraz.at}\and
lab10 collective, Graz, Austria \\
\email{p.grassberger@student.tugraz.at} \and
St.\ Pölten University of Applied Sciences, St.\ Pölten, Austria \\
\email{lukas.daniel.klausner@fhstp.ac.at}
}

\maketitle
\setcounter{footnote}{0}


% \renewcommand{\shortauthors}{More and Abraham, et al.}

\begin{abstract}
{

%\todo{rework focus to better highligh transformation}

%\sloppy
%\fussy

% Diploma fraud is a serious issue in education systems around the world.
% In addition to credentials which are forged or altered by dishonest students, there are bogus universities which only exist to make money by selling education credentials.
% It is a nontrivial problem to distinguish such bogus universities from legitimate institutions.
% On a technical level, credentials are being issued in a machine-readable format.
% The JSON schema used to encode diploma data is often unknown to the \verifier, resulting in the \verifier's inability to actually verify credential content -- an additional obstacle for global academic mobility.
%
% In this paper we introduce a document verification system to assist human \verifiers by providing them with information about the legitimacy of a \issuer.
% Recognizing the lack of a global root of trust, we utilize distributed ledger technology to provide a decentralized mechanism to authenticate universities and their credentials.
% In addition, our system is able to transform credentials into a schema understood by the \verifier by using authenticated Transformation Information provided by the community.

Recruiters and institutions around the world struggle with the verification of diplomas issued in a diverse and global education setting.
% They face challenges when interpreting diplomas issued using an unknown data schema and are concerned about diploma fraud, often requiring manual verification work.
%Not only do dishonest applicants forge or alter credentials, but there is also a whole industry of bogus institutions selling education credentials.
%Distinguishing such bogus institutes from legitimate entities is a nontrivial problem:
%
%A receiver of education credentials therefor needs to verify the credentials' trustworthiness in a tedious manual process. % to ensure that an applicant is actually sufficiently qualified.
Firstly, it is a nontrivial problem to identify bogus institutions selling education credentials.
While institutions are often accredited by qualified authorities on a regional level, there is no global authority fulfilling this task.
Secondly, many different data schemas are used to encode education credentials, which represents a considerable challenge to automated processing.
Consequently, significant manual effort is required to verify credentials.

In this paper, we tackle these challenges by introducing a decentralized and open system to automatically verify the legitimacy of issuers and interpret credentials in unknown schemas.
We do so by enabling participants to publish \transInfo, which enables verifiers to transform credentials into their preferred schema.
Due to the lack of a global root of trust, we utilize a distributed ledger to build a decentralized web of trust, which verifiers can query to gather information on the trustworthiness of issuing institutions and to establish trust in \transInfo.
Going beyond diploma fraud, our system can be generalized to tackle the generalized problem for other domains lacking a root of trust and agreements on data schemas. %similar issues in other domains without a global root of trust.
}

\end{abstract}

% \nocite{*}

%%%%%%%%%%%%%%%%%%%%%%%%%%%%%%%%%%%%%%%%%%%%%%%%%%%%%%%%%%%%%%%%%%%%%%%%%%%%%%%%%%%%%%%%%%%%%%%%%%%%
%%%%%%%%%%%%%%%%%%%%%%%%%%%%%%%%%%%%%%%%%%%%%%%%%%%%%%%%%%%%%%%%%%%%%%%%%%%%%%%%%%%%%%%%%%%%%%%%%%%%
%% !!!!!!!!!!!!!!
% https://ifipsec.org/2021/submission.html
% Submissions should be at most 14 pages long in total including references and appendices.
% Submissions should not be anonymised.
%% !!!!!!!!!!!!!!

%%%%%%%%%%%%%%%%%%%%%%%%%%%%%%%%%%%%%%%%%%%%%%%%%%%%%%%%%%%%%%%%%%%%%%%%%%%%%%%%%%%%%%%%%%%%%%%%%%%%
%%%%%%%%%%%%%%%%%%%%%%%%%%%%%%%%%%%%%%%%%%%%%%%%%%%%%%%%%%%%%%%%%%%%%%%%%%%%%%%%%%%%%%%%%%%%%%%%%%%%
{ % begin intro

\section{Introduction}
\label{sec:intro}
% intro … shorter
%
% * context, motivation, contribution
% * maybe background
% * maybe some relwork/existing work we build on?
% * → 1.5 pages

When applying for a job or a study program, applicants need to provide evidence of their past education achievements (e.\,g.\ graduation diplomas) so that the human resources department or institution is able to assess the applicants' qualifications and eligibility.
Because applicants occasionally create fake education credentials or modify credentials of legitimate issuers~\cite{Borresen2020}, universities spend considerable effort to verify the authenticity of documents.
Nowadays, education credentials can be represented as signed digital documents~\cite{EuropeanCommission2020}, e.\,g.\ as Verifiable Credentials in the Self-Sovereign Identity (SSI) context~\cite{VCETF2020}.
While digital signatures simplify the verification process, two challenges remain, which we tackle in this work -- namely to evaluate a credential's issuer's legitimacy as well as to interpret the various data formats of credentials (cf.\ \Cref{sec:concept}).

\textbf{Challenge 1:}
Because education diplomas provide substantial value, a whole industry of fake universities (so-called diploma mills~\cite{Ezell2012}) has arisen, which exist for the sole purpose of selling fake diplomas.
Within the UK alone, 240 universities were identified as such fake universities according to the UK's Department of Education~\cite{UDE2020}.
To avoid accepting such fake credentials, a verifier needs to spend considerable effort, money and resources to perform authorization of credential issuers, which is especially complex in a global setting with many different types of issuers alongside universities.

\textbf{Challenge 2:}
Besides verifying an education credential's authenticity and legitimacy, a \verifier also needs to be able to interpret the credential's content.
Various data formats have been proposed for digital education credentials~\cite{VCETF2020,HEIDI2020}, but so far, no unified, widely accepted standards have emerged~\cite{HEIDI2020}.
Therefore, verifiers have to spend additional resources to interpret the content of education credentials manually or with external help, which also hinders automated processing.

{\textbf{Contribution 1:}  \textit{Verify Legitimacy of Diverse Credentials and Issuers.}}
In this paper, we use a distributed ledger (DL) to enable all involved institutions to publish trust information about \issuers.
This effectively forms a web of trust (WoT), a directed graph of \certifications published by institutions and based on previous manual evaluations (cf.\ \Cref{sec:authz}).
% decisions of other trusted institutions.
%A web of trust is a directed graph of trust information, and institutions are able to publish outgoing edges from their own vertex, thus adding information about whom they trust to be legitimate~-- or, conversely, whom they believe to be illegitimate, if the edge carries a negative trust statement.
%Since an edge can also carry a negative trust statement, it is important that no entity has the power to delete or hide an edge, which we achieve by using a decentralized data structure (e.\,g.\ maintained by a smart contract).
This web of trust further enables \verifiers to define their own trust policies on how to automatically evaluate a (previously unknown) institution's legitimacy.

{\textbf{Contribution 2:} \textit{Verify Credentials Issued in Different Schemas.}}
Building on the above web of trust, we introduce a second graph in which involved parties can publish, discover and authenticate information to transform credentials between schemas (cf.\ \Cref{sec:transformation}).
Given this information, \verifiers are able to locally transform a \credential from the issuer's schema into their preferred schema, possibly via several intermediate schemas.
Combined with our web of trust, \verifiers can interpret a \credential's content and assess the trustworthiness of \transInfo.

Since the tackled challenges are not limited to education credentials, but also apply in other contexts, we point out that our approach can also be applied in other domains where no globally trusted authority exists.

We demonstrate the feasibility of our framework with an \textbf{implementation}, which uses a Solidity smart contract on an Ethereum-based DL to manage the web of trust and the distributed file system IPFS to store the \transInfo (cf.\ \Cref{sec:impl}).
We also evaluate and discuss several properties of our system (cf.\ \Cref{sec:discuss}).

%\textit{Beyond Education:}

} % end intro
%%%%%%%%%%%%%%%%%%%%%%%%%%%%%%%%%%%%%%%%%%%%%%%%%%%%%%%%%%%%%%%%%%%%%%%%%%%%%%%%%%%%%%%%%%%%%%%%%%%%
%%%%%%%%%%%%%%%%%%%%%%%%%%%%%%%%%%%%%%%%%%%%%%%%%%%%%%%%%%%%%%%%%%%%%%%%%%%%%%%%%%%%%%%%%%%%%%%%%%%%

%%%%%%%%%%%%%%%%%%%%%%%%%%%%%%%%%%%%%%%%%%%%%%%%%%%%%%%%%%%%%%%%%%%%%%%%%%%%%%%%%%%%%%%%%%%%%%%%%%%%
%%%%%%%%%%%%%%%%%%%%%%%%%%%%%%%%%%%%%%%%%%%%%%%%%%%%%%%%%%%%%%%%%%%%%%%%%%%%%%%%%%%%%%%%%%%%%%%%%%%%
\section{Background}% and Related Work}
\label{sec:prelims}
{ % beginning of background section

% background … short
%
% * or merge with intro?
% * → 0.5 pages, maybe shorter

In this section, we introduce concepts relevant for our approach.
%
%\paragraph{Distributed Ledger (DL)}
% \cite{Jannes2019,Alexopoulos2017}
% \cite{DBLP:journals/comsur/XiaoZLH20}
%
A \textbf{Distributed Ledger (DL)} is a redundant data store on distributed nodes without central control.
The nodes agree on a state by running a consensus protocol, and such nodes can be run by various parties, such as organizations or companies. %The nodes generally distrust each other and follow a consensus protocol to agree on a state.
%Distribution can be geographical, political, institutional, etc.\ to prevent collusion and improve resilience.
%It is more commonly known as the ``blockchain'' and can have different access models, depending on who is allowed to read from (public vs.\ private) and write to (permissioned vs.\ permissionless) the ledger.
DLs can be accessed through different access models, depending on who is allowed to participate in the network by joining it
%as user or node hosting party
(public vs.\ private) or by who has read and write access to it (permissioned vs.\ permissionless).
%
%\paragraph{Smart Contract (SC)} % (or chaincode (HL Fabric), Bitcoin scripting)
%
Many ledgers, such as Ethereum, support \textbf{Smart Contracts (SCs)}. SCs are code stored on the DL that is deterministically executed by the nodes performing the consensus protocol.
Variables in the code are stored on the DL, as well, and can be read and modified using functions supplied by the contract.
When a user sends a function call to a contract, the resulting state is only written to the ledger if all nodes agree on the result of the computation.
%Data is only stored on the DL if all nodes come to the same result when executing a contract call, thus ensuring consensus.
%We use a SC as an interface for storing and retrieving data on a DL, removing the need for a trusted (centralized) third party. %too much detail for the background
%
%\paragraph{Self-Sovereign Identity (SSI)}
%
In contrast to centralized and federated identity models~\cite{Zwattendorfer2014}, \textbf{Self-Sovereign Identity (SSI)} gives control over the identity solely to the user~\cite{Muehle2018} without having a central trusted party.
% This also removes risk of becoming the victim of a cyber attack since there is no central place storing a lot of interesting data~\cite{Fritsch2020}.
In this model, users create their own identity by using a \textbf{Decentralized Identifier (DID)}~\cite{Holt2020}.
This DID can be registered on a DL and is used to resolve key material and other information needed to interact with its subject.
In addition, it is used to link statements about the user to their identity.
%
%In SSI, every user can issue a certification of attributes of other users.
%For example, a university can certify that a student graduated from a certain study program, or an eBay user might assert that a seller was reliable.
%TODO rewrite the following sentence
%These certifications follow the W3C Verifiable Credential Data Model specification~\cite{Noble2019} and are called \textbf{Verifiable Credentials (VCs)}.
\textbf{Verifiable Credentials (VCs)} are used as a typical data format in SSI systems and are specified by the W3C~\cite{Noble2019}.
%The specification defines a generic envelope to represent a set of claims (\textit{credentialSubject}) signed by its issuer.%, as shown by example in \Cref{lst:vc1}.
% It defines a JSON structure for the credential envelope itself, while allowing credential issuers to chose their own schema for the credential subject.
While the schema of the envelope is defined by the W3C, the claims inside a VC are encoded according to a JSON schema~\cite{JSONSCHEMA} so that credentials are sufficiently flexible for various use cases.
Obviously, this flexibility of the content's schema also represents a challenge for verifiers when facing content in an unknown schema.
% \subsection{Electronic Credentials}
%The Verifiable Credential (VC) Data Model~\cite{Noble2019} is a W3C standard that defines a generic envelope to represent a set of claims about a subject in a verifiable way using digital signatures or zero-knowledge proofs.
%It defines a JSON structure for the credential envelope itself, while allowing credential issuers to choose their own schema for the credential subject.
%VCs are a central concept in the self-sovereign identity world.
%
%\lstinputlisting[caption={Example verifiable credential (proof omitted) with custom schema for the %\textit{credentialSubject}.}, label={lst:vc1}, language=json]{s/vc1.json}
% \paragraph{JSON and JSON-LD Schemas}
%
% JSON:        \cite{Bray2017}
% JSON-LD:     \cite{Kellogg2020}
% JSON schema: \cite{Wright2019}
% TODO: explain https://tools.ietf.org/html/draft-handrews-json-schema-02
% TODO: cite https://www.w3.org/TR/vc-data-model/#data-schemas → credentialSchema
%\paragraph{InterPlanetary File System (IPFS)}
%
The \textbf{InterPlanetary File System (IPFS)}~\cite{IPFS} is a distributed file system based on an open peer-to-peer network.
Similar to a public DL, there is no central party controlling the system and everyone can join the network of nodes and retrieve data.
As files on IPFS are content-addressed (i.\,e.\ a file is addressed based on a hash of its content), the files' integrity is ensured by design, which reduces the trust requirements towards the nodes.
Hence, IPFS is an ideal data storage system for decentralized apps hosted on a DL.% since storing data on a DL is costly.% or even impossible.

% \paragraph{Web of Trust (WoT)}

% \paragraph{Education Evidence}

} % end of background section

%%%%%%%%%%%%%%%%%%%%%%%%%%%%%%%%%%%%%%%%%%%%%%%%%%%%%%%%%%%%%%%%%%%%%%%%%%%%%%%%%%%%%%%%%%%%%%%%%%%%
%%%%%%%%%%%%%%%%%%%%%%%%%%%%%%%%%%%%%%%%%%%%%%%%%%%%%%%%%%%%%%%%%%%%%%%%%%%%%%%%%%%%%%%%%%%%%%%%%%%%
{% begin relwork
\section{Related Work}
\label{sec:relwork}
% relwork … short, focus on contribution, comparission
%
% * → 1 page

%\todo{max 1 page}

%\todo{comparison + less focus on edu, more on tm and credential stuff/ToIP, VC, Hyperledger}
%\todo{compare with more existing work/reviewer lit}

%\todoi{more comparison?}
%\textbf{Related Work:}
To authorize issuers of paper-based education credentials, there are only very fragmented and country- or even sector-specific accreditation lists by authorities on legitimate issuers, which moreover rarely include information for automated verification, such as public verification keys~\cite{hedd,OfsRegister,eter}.

% UDE2020: https://hedd.ac.uk/about
% OfficeStudents2018: https://www.officeforstudents.org.uk/advice-and-guidance/the-register/the-ofs-register
% ETERConsortium2020: https://www.eter-project.com

A number of projects~\cite{HEIDI2020,VCEDmodels2020,EBSIusecases2020} from public institutions and industry tackle the field of digital education credentials, with a strong focus on how to represent learning achievements.
Some projects leave the assessment of the legitimacy of the credential's issuers to the verifier~\cite{VCETF2020,Schmidt2016,IGLC2020}, while others propose centralized accreditation approaches~\cite{EuropeanCommission2020,DCC2020,Graether2018,FutureTrustConsortium2017}, often limited to single countries or the EU and focusing merely on universities.
In contrast, our approach is based on an open system, supporting the diverse needs of a global, growing and heterogeneous education world.

% HEIDI2020: cited multiple times
% VCEDmodels2020: https://w3c-ccg.github.io/vc-ed-models/
% EBSIusecases2020: https://ec.europa.eu/cefdigital/wiki/display/CEFDIGITALEBSI/Use+Cases+and+Functional+Documentation

% VCETF2020: cited pultiple times
% Schmidt2016: https://www.blockcerts.org
% IGLC2020: https://www.imsglobal.org/sites/default/files/Badges/OBv2p0Final/index.html

% EuropeanCommission2020: https://ec.europa.eu/futurium/en/europass/europass-digital-credentials-infrastructure
% DCC2020: https://digitalcredentials.mit.edu/wp-content/uploads/2020/02/white-paper-building-digital-credential-infrastructure-future.pdf
% Graether2018: cited multiple times
% FutureTrustConsortium2017: https://pilots.futuretrust.eu/gtsl

None of these projects are concerned with interoperability between different credential schemas.
Although the W3C's VC standard~\cite{Noble2019} proposes a mechanism that would allow the transformation of a credential into another representation by its subject (using zero-knowledge proofs), we are not aware of any system that enables or supports the authenticated transformation of credentials between schemas directly by the verifier.

%\todoi{DBLP:conf/csfw/LeeY10  Towards Quantitative Analysis of  Proofs of Authorization: Applications, Framework, and Techniques}

%\todoi{DBLP:journals/jcs/LiWM03  Distributed Credential Chain Discovery in Trust Management}

%\todoi{DBLP:journals/csm/DavieGHJOR19 Trust over IP (ToIP) https://trustoverip.org/}

Research literature discusses further approaches for global identity and trust frameworks~\cite{DBLP:conf/csfw/LeeY10,DBLP:journals/jcs/LiWM03}, with several publications focusing on DL technology:
Alen et al.~\cite{Allen2015} introduce the concept of a decentralized public key infrastructure (DPKI).
The concept proposes using a DL to map identifiers to key material, but does not define how to map entity identities to such identifiers.
By suggesting central registrars, it effectively reuses the concept of a certificate authority.
Trust over IP~\cite{DBLP:journals/csm/DavieGHJOR19} combines DIDs, VCs and related technology and governance models into a stack compatible with our system.
Alexopoulos et al.~\cite{Alexopoulos2017} discuss DL-based authentication and compare it to PKI and PGP's web of trust.
Yakubov et al.~\cite{Yakubov2018a} extend OpenPGP keys with the Ethereum address of their owner and provide a decentralized key server using smart contracts.
Brunner et al.~\cite{Brunner2020} provide a comparison of DL-based authentication frameworks, while Kuperberg \cite{Kuperberg2019} surveys blockchain-based identity management.

%While the frameworks described above deal with the identity of an entity, no system so far deals with the legitimacy of an entity as an education credential issuer.

} % end relwork

%%%%%%%%%%%%%%%%%%%%%%%%%%%%%%%%%%%%%%%%%%%%%%%%%%%%%%%%%%%%%%%%%%%%%%%%%%%%%%%%%%%%%%%%%%%%%%%%%%%%
%%%%%%%%%%%%%%%%%%%%%%%%%%%%%%%%%%%%%%%%%%%%%%%%%%%%%%%%%%%%%%%%%%%%%%%%%%%%%%%%%%%%%%%%%%%%%%%%%%%%
\section{Architecture Overview}
\label{sec:concept}
% solution overview …
%
% * actors, high level processes (phases)
% * Adversary Model?
% * → 1.5 pages

%\todo{max 2.5 pages}
%\todo{selling: make clear transformations are authenticated using wot}

%\subsection{Actors}

%\todo{introduce actors in less detail?}

%\todo{actors or entities?}

\begin{figure}[hb]
    \centering
    \includegraphics[width=\linewidth]{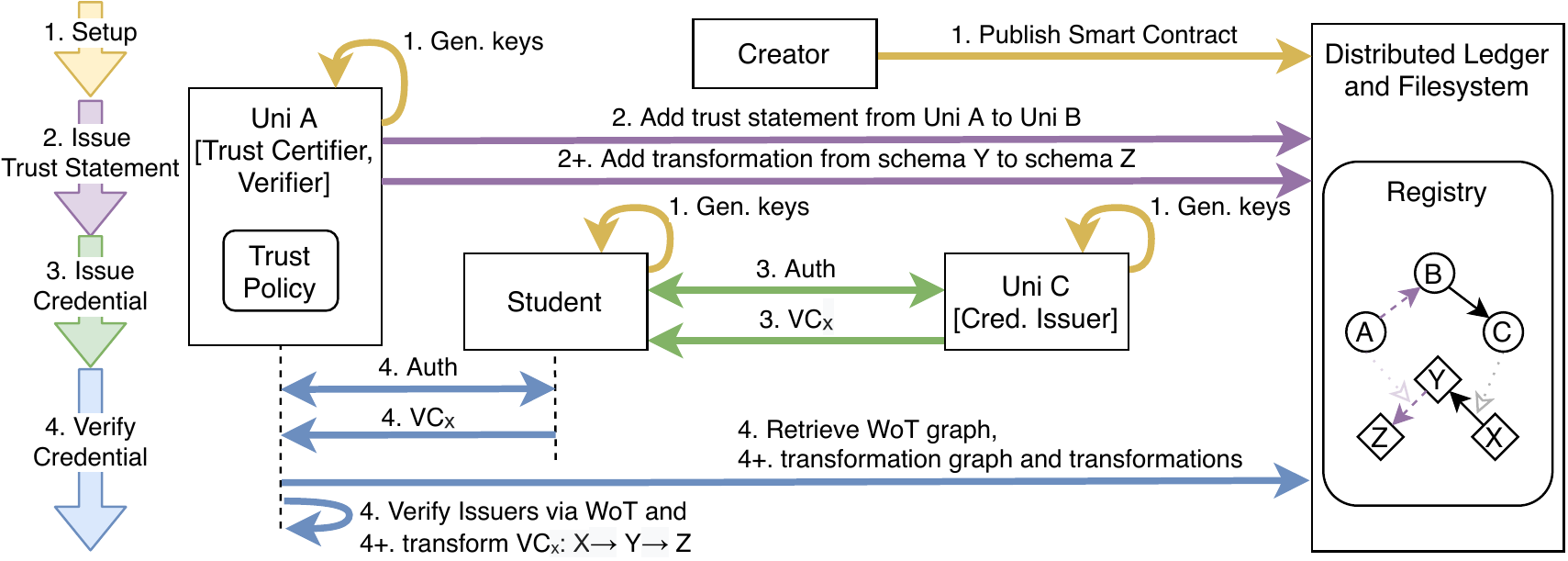}
    \caption{Architecture and example dataflows of our system. University A, acting as \certifier, adds an edge stating its trust about University B and publishes \transInfo from schema Y to its schema Z. Later, University A, acting as \verifier, verifies a credential issued by University C by additionally obtaining a previously registered edge from university B to C and a path from schema X to Y. They make sure that all \transInfo is trusted by authenticating its issuer using the WoT.}
    \label{fig:arch1}
\end{figure}

There are several \textit{actors} in our system, also shown in \Cref{fig:arch1}:

\begin{description}
    \item[\capitalStudent.]
        A natural person who earns \credentials by studying and (eventually) graduating.
        The \student is the data subject of a \credential and thus needs a key pair and a corresponding DID registered on the ledger.
        \capitalStudent{}s may later act as applicants and present their \credentials.

    \item[\capitalIssuer.]
         Any institution that issues education \credentials, e.\,g.\ a university, school, online learning provider or other training institution.
         A \issuer has a key pair and a corresponding DID registered on the ledger.

    \item[\capitalCertifier{}s.]
        All users who publish \certificates about the legitimacy (or illegitimacy) of issuers, the authenticity of \transInfo, and their trust in other \certifiers who publish \certificates.
        Credential issuers, \verifiers and any other stakeholders can also act as \certifiers{} -- enabling an open system to support a diverse education landscape.

    \item[\capitalVerifier.]
        An institution that accepts education \credentials and wants to verify them, such as an employer or university.
        A \verifier can also be a \issuer (e.\,g.\ a university is usually both).

    \item[\capitalCreator.]
        The actor that creates and publishes the smart contract of our system, which maintains a web of trust on the DL.
        It does not matter who publishes the contract, as the contract's code is public and everyone is able to verify its behavior.
\end{description}

%
% \paragraph{\capitalDL}
%A distributed ledger, formed by a set of nodes that give all users write access.
%The nodes' governance must be split among sufficiently many independent organizations such that write operations always require consensus of all nodes.
%
%\paragraph{\capitalCreator}

%\subsection{Data Items}

%The mentioned actors interact with several data items:

%\paragraph{\capitalCredential}

The actors deal with \textbf{\capitalCredentials} in W3C VC format.
The education data inside the VC envelope is encoded following a schema chosen by the \issuer.
While the schema is stated in the \credential, its interpretation is not necessarily known to the \verifier.

%\paragraph{Credential Schema}
%A credential schema is a (possibly nested) ordered list of attributes and characteristics defining a standard for describing credentials.

%\paragraph{\capitalCertificate}
Entities in our system provide \textbf{\capitalCertificates}, statements on the trustworthiness and legitimacy of others to help \verifiers in their assessment process.
\begin{definition}[\capitalCertificate]
    A \certificate is
    a tuple of the form $\langle C, \level, \confidence, \context, \uri, \sigma, \timeSymbol, I \rangle$, where
    $C$ is the \certifier,
    $I$ is the \issuer being certified,
    $\level \in [-1,1]$ is the level of legitimacy $C$ asserts for $I$,
    $\confidence \in [-1,1]$ is the level of confidence $C$ has in the statements published by $I$,
    $\context$ is the context of the certification (\textit{credential} or \textit{transformation}),
    $\uri$ is the identifier of what $I$ is trusted to issue (type of \textit{credential} or \textit{transformation}) and
    $\timeSymbol$ is a timestamp.
    Additionally, $\sigma$ is a cryptographic assurance (digital signature) of the certificate's integrity and authenticity.
    Depending on the system chosen, levels are allowed to be any real value in the interval, one of several discrete values, or even just in $\{-1,0,1\}$ (with $1$ denoting the highest trust level).
\end{definition}

%\paragraph{Web of Trust}
The \textbf{Web of Trust} is a collection of \certificates.
\begin{definition}[Web of Trust]
A web of trust is a directed, edge-labeled multigraph $W = \langle V, E \rangle$, where the vertices $V$ are entities (e.\,g.\ educational institutions) and the edges $E$ represent \certificates, which are consequently labeled as stated in the definition above.
For edge labels in a web of trust, the \certifier--\issuer parameters have to match the edge's vertices, i.\,e.\ for each edge $e$ from $v_1$ to $v_2$, the corresponding label of $e$ has to be of the form $\langle v_1, \level, \confidence, \context, \uri, \sigma, \timeSymbol, v_2 \rangle$.
%An example WoT graph is shown in \Cref{fig:graph1}.
\end{definition}

%\paragraph{\capitalTransInfo}
To facilitate interoperability between credential schemas, \certifiers also issue \textbf{\capitalTransInfo}.
\begin{definition}[\capitalTransInfo]
    Transformation information is a set of (machine-readable) transformation rules that define how to transform a \credential issued in one credential schema into another schema. \capitalTransInfo is encoded in an implementation-specific format.
\end{definition}

%\paragraph{\capitalTransInfo Graph}
Analogous to the web of trust, a \textbf{\capitalTransInfo Graph} is a collection of \transInfo.
\begin{definition}[\capitalTransInfo Graph]
    A transformation information graph is a directed, edge-labeled multigraph $TIG = \langle V, E \rangle$, where the vertices $V$ are URIs of credential schemas and each edge $e \in E$ from vertex $X$ to vertex $Y$ represents a transformation for transforming a credential of schema X into a credential of schema Y.
    %Since \transInfo itself is stored on IPFS, the edge just holds the address of the information it represents.
    An edge also contains the DID of the entity that published the \transInfo and is signed by it.
\end{definition}

%\paragraph{Registry}
The WoT and the \transInfo graph are maintained by a \textbf{Registry}, a data structure on the DL.
The registry is managed by a smart contract (SC) which provides a single point of contact for \certifiers and \verifiers who want to access or append to either graph by calling the \textit{get} or \textit{add} function on the contract, respectively.
%Edges contain information about the attestation of legitimacy and information about the transformation of the \credential between different schemas.
%The smart contract enforces some access control mechanisms for add requests, ensuring that users can only add edges outgoing from their own graph vertex.
%Since all edges in the graphs are signed by their respective \certifier, required trust in the SC itself is limited.
In addition, the registry utilizes IPFS to store the \transInfo.
While the \transInfo itself is stored on IPFS, the respective edge in the transformation graph contains the IPFS address of the information and information about its \certifier.
An example setup is shown in \Cref{fig:graph0}.

%\paragraph{Trust Policy}
A \textbf{Trust Policy} is defined by the \verifier with rules on how to decide a \issuer's legitimacy based on information obtained from the registry, i.\,e.\ trust statements and transformation information.
Such a policy also defines suitable parameters for algorithms that find paths in the graphs and calculate legitimacy scores.
%This policy defines a suitable $\pathfinder$ algorithm to find the relevant set of paths from \verifier to \issuer, and a $\calcscore$ algorithm to compute the overall legitimacy from those paths.
%The algorithms' configuration also defines whether transitive paths are allowed or only a direct path is acceptable.
%\Cref{sec:impl} discusses the algorithms in more detail.
%
%If there are no outgoing edges from the \verifier's vertex, the policy can also contain a list of vertices whose \certificates are trusted by the \verifier.
%This is effectively an unpublished edge which is only visible to the \verifier.
Additionally, the trust policy contains the rules which are used to further check the \credential's content, such as required study subjects or minimum grade point averages.
The policy is defined locally by one or more domain experts.
This approach allows \verifiers to enforce their own rules and follow their locally relevant regulations.

%%%%%%%%%%%%%%%%%%%%%%%%%%%%%%%%%%%%%%%%%%%%%%%%%%%%%%%%%%%%%%%%%%%%%%%%%%%%%%%%%%%%%%%%%%%%%%%%%%%%
%%%%%%%%%%%%%%%%%%%%%%%%%%%%%%%%%%%%%%%%%%%%%%%%%%%%%%%%%%%%%%%%%%%%%%%%%%%%%%%%%%%%%%%%%%%%%%%%%%%%
\section{Issuer Authorization}
\label{sec:authz}
% authentication architecture
%
% * → this is just WoT, so compress it to ~2 pages
% * (focus on credential issuer auth)

%\todo{maximum 2 pages}

The first step needed to check a credential is to verify the legitimacy of the credential and its issuer using information published by other entities.
This is also the basis for authenticated transformation of credentials between schemas, as described in the following \Cref{sec:transformation}.

Our approach can be split into four phases:
\begin{enumerate*}[label=(\arabic*)]
    \item Initially, the registry and its smart contract are established on the DL.
    \item The \certifier identifies a \issuer and publishes the certification information using the registry.
    \item The \issuer issues a \credential to a \student.
    \item The \student shows the \credential to a \verifier, who uses the information stored in the registry to verify the \credential and the \issuer's legitimacy.
\end{enumerate*}
The main aspects of these phases are shown in \Cref{fig:arch1} and discussed in the following paragraphs, while a formal description is given in \Cref{tab:protocolAuth}.

\begin{figure}[th!]
    \begin{mdframed}[innerleftmargin=3pt,innerrightmargin=5pt,skipbelow=0]
	\begin{description}[leftmargin=10pt]
        \setlength{\parskip}{1em}
	    \small

		\item[\pSetup:]
		\begin{itemize}[label={},leftmargin=0pt]
			\item \on{any \capitalCreator}
			\begin{enumerate*}[label={}]
				\item Generate $(\pk_\any, \sk_\any) \gets \KeyGen()$,
                \item create smart contract for WoT registry,
                \item publish smart contract at \DL
                %\item delete key pair
                \item and announce its address $\Acontract$ and code.
			\end{enumerate*}

			\item \on{all \capitalStudent{}s, \capitalIssuer{}s, and \capitalCertifier{}s}
			\begin{enumerate*}[label={}]
			    \item Generate $(\pk, \sk) \gets \KeyGen()$,
                \item create $\DID$ from $\pk$ and register $\pk$ at \DL.
			\end{enumerate*}

		\end{itemize} %end: Setup

		\item[\pIssueCert:]
		\begin{itemize}[label={},leftmargin=0pt]
			\item \on{\capitalCertifier $\Certifier$}
			\begin{enumerate*}[label={}]
			    \item Verify identity and legitimacy of issuer and obtain issuer's $\DID_\Issuer$ (out-of-band).

                \item Create the WoT edge $\edge$, which includes
                \begin{enumerate*}[label={}]
                    \item $\DID_\Certifier$,
                    \item $\DID_\Issuer$,
                    \item the level of legitimacy as issuer~$\level$,
                    \item the level of confidence as \certifier~$\confidence$,
                    \item the context $\context$ (set to \textit{credential}),
                    \item the \credential's type $\uri$ and
                    \item the current time $\timeSymbol$.
                \end{enumerate*}

                %\item create the transformation edge $\edge$, which includes
                %the Certifier's~$\DID_\Certifier$ ,
                %the certification Issuer's~$\DID_\Issuer$ ,
                %the attested level of trust in the transformation~$\level$,
                %the attested level of confidence as transformation %publisher~$\confidence$,
                %the certification's context $\context$ set to \textit{transformation},
                %and the current time t

                \item Sign $\sigma_\edge \gets \SIG.\Sign(\edge, \sk_\Certifier)$,
                \item add the signature to the edge $\edge' = \edge \| \sigma_\edge$ and
                \item publish the signed edge $\edge'$ via $\DL.\DLCall(\Acontract, \SCaddWOT, \edge')$.
			\end{enumerate*}

            \item \on{all \DL nodes}
            \begin{enumerate*}[label={}]
                \item Receive the \certifier's request and
                \item add $\edge'$ to the WoT graph's list of edges.
            \end{enumerate*}

            \item\textbf{\pRevokeCert:} as Phase \numIssueCert with different values for $\level$ and $\confidence$.

		\end{itemize} %end: Certification

		\item[\pIssueCred:]
		\begin{itemize}[label={},leftmargin=0pt]
		    \item \on{\capitalStudent $\Student$}
			\begin{enumerate*}[label={}]
			    \item Request verifiable \credential at university and send $\DID_\Student$ as well as proof of ownership of~$\sk_\Student$.
			\end{enumerate*}

			\item \on{\capitalIssuer $\Issuer$}
			\begin{enumerate*}[label={}]
				\item Verify \student's identity and ownership of~$\DID_\Student$.
                %\item verify student's ownership of~$\DID_\Student$
                \item Encode the \student's attributes~$A_\Student$ using~$\schema_\Issuer$ into $A_\Student'$ and
                \item generate \credential~$\cred$, which includes the encoded attributes~$A_\Student'$, \student's~$\DID_\Student$ and $\DID_\Issuer$.
                \item Sign the credential $\sigma_{\cred} \gets \SIG.\Sign(\cred, \sk_\Issuer)$ and
                \item issue it to the \student.
			\end{enumerate*}
		\end{itemize} %end: Issuing

		\item[\pVerify:]
		\begin{itemize}[label={},leftmargin=0pt]
			\item \on{\capitalStudent $\Student$}
			\begin{enumerate*}[label={}]
			    \item Send $\DID_\Student$, proof of ownership of $\sk_\Student$ and $\cred'$ to the \verifier.
                %\item provide $\cred$ to Verifier
			\end{enumerate*}

			\item \on{\capitalVerifier $\Verifier$}
			\begin{enumerate*}[label={}]
                \item Load up-to-date $\wot \gets \DL.\DLCall(\Acontract, \SCgetWOT)$ and issuer's $\pk_\Issuer \gets \DL.\DLResolve(\DID_\Issuer)$,
                where $\DID_\Issuer$ from $\cred'$.
                \item Verify that issuer was not revoked/expired, allowed to issue and that \credential was really signed by that issuer by verifying
                $\SIG.\Verify(\cred', \pk_\Issuer) = 1$.
                \item Find paths from \verifier to issuer relevant w.\,r.\,t.\ $\cred'$ by
                 $\paths \gets \pathfinder(\wot, \DID_\Verifier, \DID_{\Issuer}, \context_{\cred'}, \uri_{\cred'})$.
                \item Authenticate edges in $\paths$ by verifying
                $\forall \vpath \in \paths, \forall \vedge~\edge \in \vpath: \SIG.\Verify(\edge, \pk_\Certifier) = 1$,
                % $\land  \pk_{\Certifier_\vedge} = \DL.\DLResolve(\cred_{\DID_\Issuer})$
                where $\DID_\Certifier$ from $\vedge~\edge$ and $\pk_\Certifier \gets \DL.\DLResolve(\DID_\Certifier)$.
                \item Compute overall legitimacy of issuer and verify using \policy that $\calcscore(\paths) \geq \policy_{\minscore}$.
			\end{enumerate*}
		\end{itemize}%end: Verification

	\end{description}
	\end{mdframed}
	\caption{Issuer authorization protocol.}
	\label{tab:protocolAuth}
\end{figure}

\textbf{\pSetup:}
To set up the system, the \creator publishes a smart contract on the DL which maintains the registry.
This contract provides a single point to create or update ($\SCadd$) and retrieve ($\SCget$) the lists of edges.
%The \creator then announces the contract's address and code to the public.
%This setup phase only has to be performed once in the lifetime of the registry.
Additionally, \students, \issuers and \certifiers create their own key pairs and use them to establish their self-sovereign identities, i.\,e.\ derive and register DIDs.
%\capitalVerifier{}s are not required to have a DID.
These DIDs with corresponding DID documents are registered on the DL to enable other parties to retrieve the respective public keys for signature verification processes.

\textbf{\pIssueCert:}
After having assessed a \issuer's legitimacy (e.\,g.\ in a previously performed tedious manual process or using other channels), a \certifier may share its decision with others by issuing a \certificate.
% This assessment could also stem from the verification of previous \credentials by the same \issuer.
Based on the previous assessment, the \certifier chooses an appropriate level of legitimacy $\level$, which may also be a negative value if the \certifier arrives at the conclusion that the \issuer is an illegitimate institution.
In addition, it adds the level of confidence in \certificates published by the \issuer, represented by $\confidence$.
The \certifier publishes its assessment as a \certificate on the registry
%Besides the DIDs of \certifier, the target (e.\,g.\ \issuer or \certifier), and the level of legitimacy, the edge also contains additional data:
%The context $\context$ is set to \textit{credential} and an identifier $\uri$ defines the credential's type (e.\,g.\ \textit{Bachelor's degree}).
%If the \certifier trusts the certified \issuer to also issue \certificates on their own, it says so by adding this confidence level $\confidence$ to the edge.
%Also, the current time is added to the edge to allow filtering for certifications that were valid at a certain point in time.
by setting the variables on the edge accordingly (cf.\ \Cref{sec:concept}), using its private key to sign the edge, and publishes the signed edge by calling the smart contract through one of the DL's nodes.
This creates a new edge in the registry's WoT graph pointing from the \certifier to 
%possibly intermediate \certifiers to 
the prospective \issuer.
%
 % explain the attested level of legitimacy as university~$\level$,
 % TODO: explain the attested level of confidence as transformation publisher~$\confidence$,
%
 % \todoi{explain that \verifier needs to certify some unis on their own, or trust them by other means → as starting point for trust path}
 % (a certification is not valid if there is no path from the \verifier to it; otherwise fake unis could just certify themselves)
%
%The DL node adds the call to the ledger, which triggers execution of the smart contract's $\SCaddWOT$ function on the other nodes.
%The contract checks that the entity making the call is set as \certifier in the edge, i.\,e.\ the edge is outgoing from the \certifier's vertex.
%If this is the case, and the DL nodes reach a consensus, the new edge is added to the graph, and the graph is updated in the ledger.
Optionally, the \certifier may also issue \transInfo or trust other entities to do so (cf.\ \Cref{sec:transformation}).

\textbf{\pRevokeCert:}
As trust in other institutions changes and faulty entries occur,
%\certifiers require functionality to revoke or adapt their previously issued \certificates.
%in our system
\certifiers are able to add a new edge between the same vertices (but with a more recent timestamp) to the WoT, thereby altering the stated level of legitimacy or other attributes.
Based on the timestamps, \verifiers select edges at points in time based on the received \credential and their trust policy.

\textbf{\pIssueCred:}
After completing a course or graduating from a university, the \student requests
%that the respective institution issues
a VC for this accomplishment.
Initially, the \student proves their identity using their DID and private key.
%e.\,g.\ by using their private key to sign the \issuer's challenge, which the \issuer then verifies with the \student's public key from the DID document stored in the DL.
Then, the \issuer loads the \student's attributes from its \student information system, encodes those attributes using a suitable JSON schema and places the encoded attributes into a VC.
The \issuer finally signs the VC and hands it to the requesting \student for their own use.
%, e.\,g.\ to present in a job application process.
%
The issuing of \credentials is independent of the registry and a DL may only be needed to retrieve the \student's public key via its DID for authentication.
%A DL might also be used to register the existence of the VC for revocation status checks.

\textbf{\pVerify:}
When a \student presents a \credential to a \verifier (e.\,g.\ within an application process), the \verifier needs to verify it according to its trust policy.
The \verifier first retrieves the registry's WoT graph by calling the smart contract's $\SCgetWOT$ function using a DL node.
On this graph, the \verifier performs the path-finding algorithm defined in its policy.
%This algorithm identifies all trust paths (i.\,e.\ lists of \certificates) from a set of trusted vertices to the \issuer's vertex.
The \verifier also verifies the signature of each \certificate in the trust paths with public keys obtained from the DL.
If there are multiple paths towards a \issuer, the \verifier's trust policy may select which paths are relevant.
This enables different verification scenarios, such as ``Was the \issuer trusted at the point in time the \credential was issued?'' and ``Is the \issuer trusted right now?''.
On the identified paths, the \verifier computes an overall ``legitimacy score'' and compares this score with the policy's requirements.
Next, the \verifier verifies the \credential's signature with the public key that has been registered in the DL for the \issuer's DID, including a revocation check by consulting the respective revocation registry.
%Since that \credential may have been revoked,
The \verifier concludes the process by checking the \credential with regard to a set of rules stated in the trust policy, such as the precise field of study or certain grade requirements.
Optionally, if the \verifier does not support the schema of the received \credential, the \verifier may engage in a transformation process and transform the \credential into a supported schema as described in \Cref{sec:transformation}.
%In short, the \verifier obtains \transInfo from the DL and uses it to transform the \credential, possibly over multiple steps, into a supported schema.

%%%%%%%%%%%%%%%%%%%%%%%%%%%%%%%%%%%%%%%%%%%%%%%%%%%%%%%%%%%%%%%%%%%%%%%%%%%%%%%%%%%%%%%%%%%%%%%%%%%%
%%%%%%%%%%%%%%%%%%%%%%%%%%%%%%%%%%%%%%%%%%%%%%%%%%%%%%%%%%%%%%%%%%%%%%%%%%%%%%%%%%%%%%%%%%%%%%%%%%%%
\section{Credential Transformation}
\label{sec:transformation}
% transformation system
%
% * → this is the meat, ~3 pages

%\todo{maximum 2 pages}

The second step during checking of a \credential is to check the content of the \credential using a local policy.
%If the credential is not in a schema which the \verifier understands,
The transformation steps described in this section help \verifiers who are not familiar with the issued \credential's schema and, therefore, need support to interpret the \credential correctly.

\begin{figure}[t]
{
    \centering
    \subfloat[Registry setup: Edges in the Transf.\ Graph are authenticated using the WoT.\label{fig:graph0}]{%
        % The WoT graph contains \certificates which can be used to verify the legitimacy of \issuers and transformations. While the \transInfo itself is stored on IPFS, the transformation graph stores their IPFS address and signature.
       \includegraphics[width=0.45\linewidth]{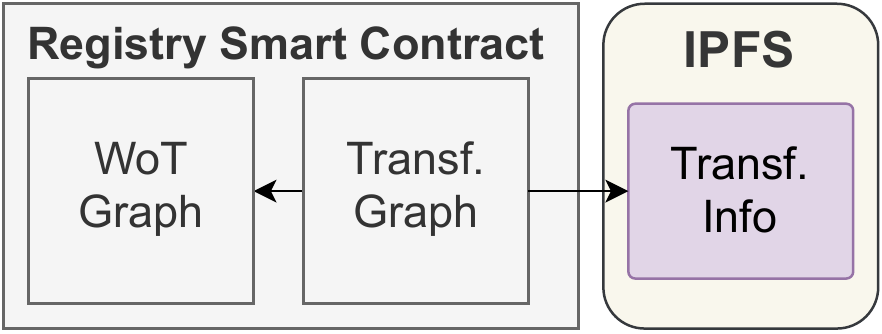}%
    }\hfil
    \subfloat[Transformation process using Transf.\ Info discovered using the Transf.\ Graph.\label{fig:transformation}]{%
       \includegraphics[width=0.45\linewidth]{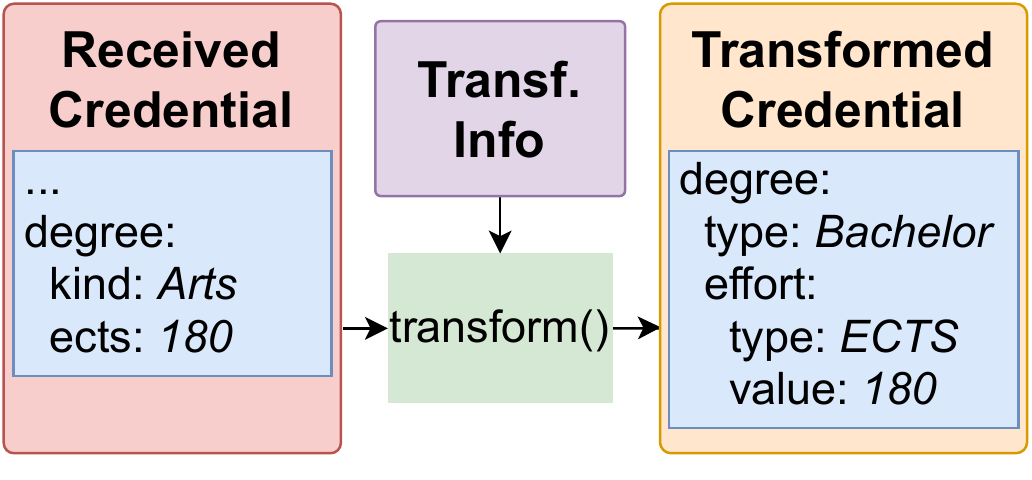}%
    }

    \caption{Example transformation with information authenticated using the WoT.}
    \label{fig:transformationFigures}
}
\end{figure}

We extend two of the previously described phases to enable transformations between schemas:
\begin{enumerate*}
    \item[\numIssueTrans{}] \capitalCertifiers knowing how to interpret a \credential's schema and transform it into another schema publish this \transInfo to the DL's registry to help others.
    For example, a university having dealt with \credentials in a foreign schema previously (as \verifier) may publish its \transInfo into a local schema (as \certifier).
    To establish trust in this \transInfo, we build upon the web of trust approach as established in phase \numIssueCert above.
    \item[\numVerifyTrans{}] After verifying the \student's \credential, the \verifier
    %may need to
    transforms the \credential into a supported schema by querying the transformation graph for suitable \transInfo, verifying its authenticity and its publisher's legitimacy using the WoT graph.
\end{enumerate*}

\Cref{fig:graph0} illustrates how the WoT is used to authenticate the \transInfo, and a formal view on the extended phases is given in \Cref{tab:protocolTranf}.
%\textcolor{red}{Felix says: Text and ref for Fig 3a does not match.}

\textbf{\pIssueTrans:}
%If the \certifier already verified a \credential from that \issuer before, it is likely that it has knowledge about the schemas involved.
%In that case, the \certifier can add information about how to transform a \credential
%from the schema defined by its issuer to the one understood by the \certifier.
%between those schemas to the transformation graph.
If a \certifier has information about how to transform a credential from one schema to another, they can add this information to the transformation graph.
As the \transInfo may be too large, it is not stored on the ledger itself.
Instead, the \certifier publishes the \transInfo on IPFS, which results in a content-dependent IPFS address (basically the information's SHA256 hash).
This address is then added to the new edge and sent to the transformation graph by calling $\SCaddT$.

\textbf{\pIssueTCert:}
Analogous to the assessment of a \issuer's legitimacy, \certifiers assess and share the legitimacy to publish \transInfo and thereby state that they trust the other entity and its \transInfo.
%, which specifies a mapping between schemas.
This statement is published in the same way as in phase \numIssueCert, with the context $\context$ set to \textit{transformation}, and added to the WoT graph by calling $\SCaddWOT$.
The identifier $\uri$ defines the respective schema.

\textbf{\pVerifyTrans:}
After phase \numVerify has verified the \credential as well as the \issuer's legitimacy, the \verifier
%requires help in interpreting the \credential if the \verifier is unfamiliar with the \credential's schema.
%Therefore, the \verifier
retrieves \transInfo from the registry by calling $\SCgetT$ and uses it to transform the \credential into a schema known to the \verifier's policy engine, as illustrated in \Cref{fig:transformation}.
Initially, the \verifier loads the transformation graph and finds a path from the \credential's schema to the known schema.
In each step, while traversing the path, the \verifier verifies the signature on the edge and assesses the legitimacy of the edge's publisher using the WoT graph.
It is also possible to define a rule in the policy requiring the creator of a transformation to possess a minimum legitimacy level and discard all other transformations.
The \verifier then loads the transformation from IPFS and verifies its integrity by computing the hash value and comparing it with the (content-dependent) address value stored on the edge.
Afterwards, the \verifier executes the transformation on the \credential.
If multiple relevant paths with \transInfo exist, it depends on the trust policy which one is used -- or if all are used and a trust decision is made by a human actor.
Finally, the \verifier obtains a \credential represented in a supported schema, which simplifies further checks according to the trust policy (or manually by humans).

\begin{figure}[t!]
    \begin{mdframed}[innerleftmargin=3pt,innerrightmargin=5pt,skipbelow=0]
	\begin{description}[leftmargin=10pt]
        \setlength{\parskip}{1em}
	    \small

		\item[\pIssueTrans:]
		\begin{itemize}[label={},leftmargin=0pt]
			\item \on{\capitalCertifier $\Certifier$}
			\begin{enumerate*}[label={}]
                \item Define transformation information~$\transl_{X \rightarrow Y}$ from~$\schema_X$ to~$\schema_Y$
                and publish it on~$\IPFS$ on address~$\Atranslation$.

                \item Create transformation edge $\edge_T$, which includes
                \begin{enumerate*}[label={}]
                    \item $\DID_\Certifier$,
                    \item the URIs identifying the source schema $\schema_X$ and target schema $\schema_Y$,
                    \item the IPFS address of the transformation~$\Atranslation$ and
                    \item the current time $\timeSymbol$.
                \end{enumerate*}

                \item Sign the edge $\sigma_{\edge_{\Ttrans}} \gets \SIG.\Sign(\edge_T, \sk_\Certifier)$ and
                \item send it to contract by calling $\DL.\DLCall(\Acontract, \SCaddT, \edge_\Ttrans')$.
			\end{enumerate*}

            \item \on{all \DL nodes}
            \begin{enumerate*}[label={}]
                % \item Receive the \certifier's requests to add the edge to the transformation graph
                %\item verifiy that the caller's DID corrosponds to $\DID_\Certifier$\newline
                %in the given edge
                \item Add $\edge_\Ttrans'$ to the transformation graph's list of edges.
                %, overwriting other edges with same $(\DID_\Certifier, \DID_\Issuer)$ pair
            \end{enumerate*}

            \item\textbf{\pIssueTCert:} as Phase (2) with
            the certification's context $\context$ set to \textit{transformation}
            and the identifier $\uri$ to a URI identifying a \credential schema.
		\end{itemize}

		\item[\pVerifyTrans:]
		\begin{itemize}[label={},leftmargin=0pt]
			\item \on{\capitalVerifier $\Verifier$}
			\begin{enumerate*}[label={}]
                %\item[] \textit{(optional)} transform $\cred$ to $\schema_\Verifier$ and verify policy:

                \item Load up-to-date transformation graph $\transf \gets \DL.\DLCall(\Acontract, \SCgetT)$ and
                \item find paths from issuer's to \verifier's schema by
                 $\paths_\Ttrans \gets \pathfinder(\transf, \schema_\Issuer, \schema_\Verifier)$.

                \item For each $\vpath$ in $\paths_\Ttrans$ and each $\vedge$ in $\vpath$,
                \begin{enumerate*}[label={}]
                    \item authenticate $\vedge$ by using $\wot$,
                    %\item verify that $\confidence_\edge$ is higher than $\policy_{\minconf}$
                    \item extract transformation address~$\Atranslation$,
                    \item load transformation~$\transl \gets \IPFS.\load(\Atranslation)$ and
                    \item transform \credential~$\cred' \gets \translate(\transl, \cred)$ for the next iteration.
                \end{enumerate*}
                \item Each time, verify that $\cred$ fulfills $\policy_{\rules}$ or reject stating a reason.
                %   if not reject $\cred'$ with reason why

                \item Accept $\cred$ for further human processing.

                % \item interact with $n$ nodes to seal $\DID_\Prover$, i.e. \newline
                % $\forall i \in [n]: \sigma_i \gets \MSIG.\Sign(\DID_\Prover, \sk_i) $ and\newline
                % $\sigma_{\DID_\Prover} \gets \MSIG.\ASigs(\{(\pk_i, \sigma_i)\}_{\forall i \in [n]})$
                % \item issue $\sigma_{\DID_\Prover}$ to the user
                % \todo[inline]{also generate seal on public key of DID, so that this can be forwarded when offline}
			\end{enumerate*}
		\end{itemize} %end: Verification

	\end{description}
	\end{mdframed}
	\caption{Credential transformation protocol.}
	\label{tab:protocolTranf}
\end{figure}

Of course, transformations between schemas have limitations.
For example, if a target schema requires values that are not available in the source schema, a direct transformation might not be possible.
In that case, the \verifier or its policy have to decide whether the presented \credential is considered sufficient or whether in light of these missing values (which are e.\,g.\ simply mapped to defaults), is is deemed insufficient for the assessment.

%%%%%%%%%%%%%%%%%%%%%%%%%%%%%%%%%%%%%%%%%%%%%%%%%%%%%%%%%%%%%%%%%%%%%%%%%%%%%%%%%%%%%%%%%%%%%%%%%%%%
%%%%%%%%%%%%%%%%%%%%%%%%%%%%%%%%%%%%%%%%%%%%%%%%%%%%%%%%%%%%%%%%%%%%%%%%%%%%%%%%%%%%%%%%%%%%%%%%%%%%
\section{Prototype}
\label{sec:impl}
% prototype
%
% * eval
% * → 1 page

%\todo{shorten to 1 page, maybe merge with meat/discussion}

This section describes the prototype implementation of our concept, which highlights the feasibility of our system.

\textbf{Distributed Ledger (DL):}
We host our smart contract and its registry on an Ethereum-compliant DL.
While the choice of permission model depends on the concrete use case, we used a public permissionless ledger for the proof-of-concept implementation, thus allowing participation without prior registration.
While this is not a requirement, we host both graphs inside the same smart contract and thus on the same ledger.
%
%\textbf{Smart Contract and Registry}
The registry was implemented as a smart contract using the Solidity smart contract language~\cite{soliditylang}.
%\Cref{lst:contract1} in \Cref{app:code} shows a simplified example of a smart-contract-based registry.
%After a \certifier calls the contract's $\SCadd$ function, the contract first verifies if the caller of the contract (\textit{msg.sender}) is represented by the given \certifier DID (\textit{certiDID}).
%This is done by resolving the \certifier's DID (\textit{certiDID}) and comparing the retrieved ledger address with the sender's.
%If the addresses match, the provided edge is added to the graph.
%After a \certifier calls the contract's $\SCadd$ function, the provided edge is added to the graph.
%
%\textbf{Listener DL Node}
As an optional component, we operate a lightweight ``listener'' node inside the trust boundary of the \verifier as an optional performance optimization to reduce the latency between the \verifier's client and the DL node as well as to mitigate censorship attacks (cf.\ \Cref{sec:discuss}).
This node receives and verifies all blocks of the ledger, but does not participate in the consensus protocol.
Since the \verifier is only concerned with the state of the registry's smart contract, all other transactions can be discarded, thus minimizing storage needs.
%While this ``listener'' node is optional, it ensures integrity in the retrieved contract state and faster retrieval of the graphs for the \verifier.
%Additionally, the node enables the \verifier to request the state of the registry at a certain point of time in the past.
%If the graphs are cached, all IPFS addresses mentioned in the \transInfo graph can also be resolved and the files retrieved, making them rapidly available to the \verifier when needed.

\textbf{Distributed Storage Layer:}
As \transInfo data is potentially too large to be stored directly on the DL, our implementation uses IPFS as the distributed storage layer due to its availability and maturity, although other distributed file systems could be used instead.
Important properties are integrity protection of data and content-addressable file resolution.
%A brief discussion concerning the availability of IPFS is given in \Cref{sec:discuss}.

%The availability of files stored on IPFS is weakened by the fact that there is no guarantee that nodes serving a certain file keep the file or even stay online at all.
%To mitigate this issue, the IPFS node trusted by the \verifier pins all files mentioned in the transformation graph, thus keeping them available for the \verifier and its trusted parties (and therefore also for the IPFS network as a whole).

\textbf{Client Components:}
% TODO \cite{web3jsGithub}
% TODO \cite{MetaMask}
To add and retrieve edges from the registry, we implemented client components in JavaScript.
All components use the Ethereum JSON RPC interface~\cite{ethapi} to communicate with DL nodes.
For the \certifier, we use \textit{web3.js} to create transactions and the \textit{MetaMask} browser extension as the wallet to sign and send them to the DL.
For the \verifier, we implemented a client to retrieve and visualize the WoT graph and to compute paths in the graph, serving as the data source for the verification.
%\todoi{Peter G.: please verify/fix; or kill section? <- verified.}

\textbf{Transformation System:}
Transformation information is discovered using the registry's transformation graph and authenticated using the WoT graph, as explained in \Cref{sec:transformation}.
%An example transformation is shown in \Cref{fig:transf2}.
The \transInfo can be encoded by any means as long as the \verifier is able to execute them.
In our implementation, we used \textit{jsonpath-object-transform}~\cite{jsonpathott} and published corresponding templates encoded in JSON to IPFS.
Other approaches to transform JSON between different schemas or to another representation (such as XML) include table-based transformations, more generic templating systems using JSONPath~\cite{JsonPath} and systems based on XML's XSLT (including its experimental variant JsonT~\cite{jsont}).

% TODO \cite{Friesen2019}

\textbf{Trust Policy:}
The policy not only defines how to evaluate a \credential's legitimacy, but also specifies a set of rules on whether the content of the \credential is acceptable.
For example, a university might want to restrict acceptable credentials to Bachelor's certificates during the application process for a Master's program.
Our system does not hard-code such rules, but instead hands over all retrieved data to a policy system~\cite{DBLP:journals/jcs/BeckerFG10,DBLP:conf/ifiptm/ModersheimSWMA19,ssitpl21}.
This policy system then executes a policy defined by domain experts, such as the university's registrar's office, providing them with a high-level way to define rules~\cite{DBLP:conf/openidentity/ModersheimN19,DBLP:conf/icissp/WeinhardtO19,DBLP:conf/openidentity/WeinhardtP19}.

\vspace{-1em}

%%%%%%%%%%%%%%%%%%%%%%%%%%%%%%%%%%%%%%%%%%%%%%%%%%%%%%%%%%%%%%%%%%%%%%%%%%%%%%%%%%%%%%%%%%%%%%%%%%%%
%%%%%%%%%%%%%%%%%%%%%%%%%%%%%%%%%%%%%%%%%%%%%%%%%%%%%%%%%%%%%%%%%%%%%%%%%%%%%%%%%%%%%%%%%%%%%%%%%%%%
\section{Discussion}
\label{sec:discuss}
% discussion
%
% * → 1 page

%\todo{shorten to 1 page}

This section discusses further aspects of our concept and prototype.

%\textbf{Local Trust Decisions} %\todo{changed from trust locality, ok?}
%Every \verifier needs to be able to define for themselves 1) under which conditions a \issuer is considered legitimate and 2) which rules a \credential needs to fulfill.
%To offer this flexibility, our approach does not fix those conditions and rules, but rather enables \verifiers to define their own trust policies, which define a relevant point in time, algorithms to compute the ``legitimacy score'', and a set of constraints a \credential needs to fulfill.
%This leaves the ultimate decision of whom \verifiers consider legitimate to the \verifiers themselves.

\textbf{Local Trust:}
A tackled challenge is the authenticity and legitimacy of \credentials that use unknown schemas.
%If an external entity transforms the \credential to a supported schema before handing the transformed \credential to the \verifier, the payload obviously changes and, thus, the original signature becomes invalid.
To avoid a signature becoming invalid due to transforming the \credential, the \verifier first verifies the signature of the original \credential and only then converts the \credential (based on trusted \transInfo) locally.
Both signature verification and \credential transformation occur locally in the \verifier's trust domain, and only trusted \transInfo is used.
Consequently, it is guaranteed that the transformed \credential has the same meaning as the incoming signed one.

%do we reaLLy need this point
\textbf{Revocation:}
Revocation of \certificates is an important feature in any trust management system~\cite{Abraham2020}.
In our proposed system, a \certifier can revoke a \certificate at any time by issuing a new \certificate with a reduced legitimacy level -- even with a negative value to explicitly declare mistrust in the \issuer.
Additionally, a \issuer can revoke a \credential after it has been issued, e.\,g.\ by publishing a corresponding statement to a revocation registry, which is checked by the \verifier.
Checks for both types of revocations need to be performed during \credential verification by the \verifier.

\textbf{Smart Contract Security:}
Depending on the access model of the DL, many or even all entities can write data into the registry.
Since all information published in the registry is signed by its publisher, this is not a problem for the authenticity of \transInfo and \certificates.
 %, \issuers and their \credentials.
Nevertheless, an overfull registry could lead to performance issues during verification.
To mitigate this issue, the smart contract serving the registry can be equipped with access control mechanisms.
For example, it is possible to enforce that a \issuer can only add edges which are outgoing from its own vertex.
Preventing an attacker from adding invalid edges starting at the vertex of a legitimate entity keeps the path-finding algorithms from having to verify many invalid signatures to find the valid ones.
Furthermore, it is possible to only allow adding edges to \certifiers who are already part of the graph.
While this keeps attackers away, it also hinders legitimate entities from joining the graph.
Thus it depends on the concrete use case which mechanisms should be used.

Since a smart contract is code and code could contain bugs that malicious parties might be able to exploit~\cite{DBLP:conf/ccs/TorresBNPJM20}, it is important to apply secure software development practices when adding access control mechanisms.
This is even more important when considering that smart contracts cannot be changed after they have been deployed on a DL~\cite{DBLP:conf/ndss/RodlerLKD19,DBLP:journals/corr/abs-2010-00341}.

\textbf{Operational Costs:}
%\todo[inline]{Consider removing operational costs and only add the most important information to the paragraph DL Performance Evaluation (1-2 sentences)}
To evaluate our concept, we deployed a smart contract that maintains a registry as described in \Cref{sec:concept}, which manages the two graphs represented as a list of edges.
The smart contract does not perform additional checks to verify the publisher, so a \verifier needs to retrieve the full graphs and filter out any irrelevant or invalid edges.
While a public DL enables an open system, all write operations on the ledger have a cost.
%~\cite{Lee2020}.
In the Ethereum context, the cost of a contract call is measured in units called \textit{gas} and depends on the required computational effort.
%Adding an edge to our registry costs about 78\,000 gas.
%The price of gas in Ethereum's cryptocurrency \textit{Ether} is determined by the current load of the DL network.
% Since Ether's conversion rate to other currencies is not fixed, there are in total two levels of uncertainty when converting to e.\,g.\ US dollars.
%Ether's conversion rate to other currencies is not fixed and depends on various factors.
%First, the price of an Ether in US dollars depends on various market conditions.
%Second, the price of a gas in Ether depends on the current transaction volume of the Ethereum ledger.
%It is thus not possible to state a stable cost for our contract~\cite{DBLP:journals/cee/PoongodiSVBSIK20}, but
%At the time of writing, the price\footnote{Computed by multiplying gas price (\url{https://etherscan.io/chart/gasprice}) and ether price (\url{https://etherscan.io/chart/etherprice}) on 22 January 2021.} of adding a new edge to the graph is around US\$ 10.
Adding an edge to our registry costs about 78\,000 gas, worth about US\$~10 in January 2021.\footnote{Computed by multiplying gas price (\url{https://etherscan.io/chart/gasprice}) and ether price (\url{https://etherscan.io/chart/etherprice}) on 22~January 2021.}
In contrast to write operations, read operations are free.
As the majority of operations in our system are of the latter kind, the total costs are still relatively low.

%\todoi{also argue why this amount of gas? (size of contract, operations and algorithm)}
% ~\cite{DBLP:journals/sncs/LealCG20}
Since costs of using a public ledger are hard to predict in advance, private or consortium ledgers represent alternatives.
In such ledgers, members of a consortium operate all nodes of the ledger, removing the need for an incentive system like gas.
This limits the expenses to the costs needed to host the nodes but restricts (write) access to consortium members.

\textbf{Performance Evaluation:}
%To evaluate the performance of our implementation, we ran benchmarks on an Ethereum-based DL.
We evaluated graphs with up to 10\,000 edges and measured the performance of different operations using an Ethereum-based DL.
%~\cite{DBLP:conf/scc2/WangYMX20}.
Adding a new edge to the graph is not a time-critical operation and only performed whenever a new \certificate is issued.
The \textit{add} call to our contract only took less than 1s, while integration of the edge into the ledger depended on the ledger's block time (around 13s for mainnet Ethereum~\cite{ethblocktime}).
Likewise, the retrieval of a graph with 10\,000 edges from remote DL nodes (with \textit{get}) took
%about 100~ms for a graph with 100~edges.
%, and an experimental loading of 10\,000 edges was done in
under one second.
%\todo{1s vs 100*100=10s?}
Performance is of even smaller concern if the \verifier retrieves edges from a local ``listener'' node (cf.\ \Cref{sec:impl}).

\textbf{Sybil Attacks and Censorship:}
%\textbf{Avoiding Certification Mills (Sybil Attack)}
%\textbf{Mitigating Sybil Attacks}
%\todo{is this something this work achieves or a feature of WoT?} TODO check if we need the sybill attack
The existence of fake universities issuing (fake) \credentials also makes the existence of fake \certifiers
issuing \certificates to (fake) universities
plausible.
However, since
%While the existence of fake \certifiers is plausible,
such fake \certifiers are neither trusted by a \verifier nor have a trust path from the \verifier to them, these (fake) \certificates have no influence on a \credential's verification.
%, as shown in \Cref{fig:graph2}.
%
%\textbf{Censorship Resistance:}
%\cite{Vasek2014}
%\citet{Alexopoulos2017} provide a formal model of different authentication systems, arguing in favor of systems based on distributed ledger technology.
Although the decentralized and distributed nature of DLs provides resistance against censorship and denial-of-service attacks~\cite{Alexopoulos2017},
%Also, attacks in which a malicious \certifier provides different data to a specific \verifier than to the other participants are not possible as all participants have the same view of all trust relations at any given time~\cite{Alexopoulos2017}.
% \textit{stealth targeted attacks}~\cite{Alexopoulos2017}
%
% avoiding last mile attacks
%Although the published graphs themselves are a result of a consensus protocol and therefore agreed on by all DL nodes, a node might still provide bogus information to a \verifier.
%To mitigate this censorship on the last mile,
a node might still provide bogus information to a \verifier, hence a \verifier needs to establish a trust relationship with at least one node.
One way of doing this is having the \verifier operate its own node as discussed in \Cref{sec:impl}.
Another way is to
%use a multi-signature scheme and
ask multiple or even all nodes to attest a certain set of edges represents the full graph and that no edges were censored.
% out of scope → next paper

%The same holds for the distributed storage system.
%Depending on the requirements for availability, it is necessary that a verifier establishes a trust relationship with an IPFS node as well. %TODO is this sentence important?

%%%%%%%%%%%%%%%%%%%%%%%%%%%%%%%%%%%%%%%%%%%%%%%%%%%%%%%%%%%%%%%%%%%%%%%%%%%%%%%%%%%%%%%%%%%%%%%%%%%%
%%%%%%%%%%%%%%%%%%%%%%%%%%%%%%%%%%%%%%%%%%%%%%%%%%%%%%%%%%%%%%%%%%%%%%%%%%%%%%%%%%%%%%%%%%%%%%%%%%%%
\section{Conclusion}
\label{sec:conclusion}
% conclusion

% * 0.5 pages, maybe shorter

%\todo{adapt to new focus, maybe shorten a bit}

In this paper, we introduced a system that simplifies the time-con\-sum\-ing and costly task of verifying \credentials and the legitimacy of the \credentials' issuer.
While we placed the focus on the education domain, our approach can be generalized to other domains facing the same problems. %similar challenges.
%
%In a global and increasingly diverse education world, it is hard to decide who is allowed to issue which kinds of diploma and other credentials.
%Without such verification, bogus universities can sell fake diplomas, with possibly disastrous consequences.
%Moreover, the fragmentation of data schemas complicates checks based on the content of \credentials issued in an unknown schema.
%
Our open and decentralized system introduces a web of trust maintained by a smart contract on a DL.
We build upon this trust assessment framework for \transInfo, as well.
Participants of our system publish \transInfo on IPFS and a \certificate therefor on the DL.
If confronted with a \credential in an unknown schema, \verifiers look up a transformation chain on the DL, as well as trust paths within the WoT to assess the legitimacy of these transformation steps.
As a result, \verifiers are able to transform received \credentials to a supported data schema and automatically check their content.
%This web of trust is grown by participants of our system, who publish \certificates.
%Upon receiving a \credential, \verifiers search for paths towards the \credential's issuer within the web of trust and evaluate these paths to assess the legitimacy of the \issuer according to the \verifiers' own policies.
%Consequently, \verifiers may reuse decisions of other participants (if they are sufficiently trusted) rather than having to perform a costly and time-consuming manual assessment themselves in every single case.
%
% open, decentralized system
% maintain web of trust on DL
% entities can certify trustworthiness
% \verifiers find trust path from trusted entity to \issuer of credential
%
%Moreover, the fragmentation of data schemas complicates checks based on the content of \credentials.
%We tackle this challenge by building upon the trust assessment framework for \transInfo, as well.
%Participants of our system publish \transInfo on IPFS and a \certificate for that information on the DL.
%If confronted with a \credential in an unknown schema, \verifiers look up a chain of transformations on the DL, as well as trust paths within the web of trust to assess the legitimacy of these transformation steps.
%As a result, \verifiers are able to transform received \credentials to a supported data schema.
%
% build upon this trust assessment framework also for transfomation
% publish the transformation information on IPFS, and trust statement for that \transInfo on the web of trust
%
Finally, our {implementation} demonstrates the feasibility of our concept while the extensive {discussion} highlights additional benefits, such as the \verifier's control over their trust decisions and the ledger's resistance against censorship.

\iffalse
%from intro:

\paragraph{Benefits.} % benefits + positive impact
Our system reduces the effort to verify credentials and the legitimacy of their issuers even across different data formats.
The system offers a platform to reuse the legitimacy assessments and transformation logic of other trusted participants, but does not force any definition for those goals upon verifiers.
The contributions of our paper can complement other existing systems and improve their interoperability w.\,r.\,t.\ trust assessment and schema transformations.
Overall our system saves time, improves the quality of verification outcomes, counters diploma fraud and helps to make sure that applicants have the necessary qualifications for their desired positions.

\fi

%%%%%%%%%%%%%%%%%%%%%%%%%%%%%%%%%%%%%%%%%%%%%%%%%%%%%%%%%%%%%%%%%%%%%%%%%%%%%%%%%%%%%%%%%%%%%%%%%%%%
%%%%%%%%%%%%%%%%%%%%%%%%%%%%%%%%%%%%%%%%%%%%%%%%%%%%%%%%%%%%%%%%%%%%%%%%%%%%%%%%%%%%%%%%%%%%%%%%%%%%
% references …
%
% * maybe not 73 → edu references mergen, important links to footnotes?
% * (cleanup links, remove unnecessary info, etc.)
% * → 1.5 pages?

%\bibliographystyle{ACM-Reference-Format}
\bibliographystyle{splncs04}
\bibliography{wot}


\ifnum\thepage>16
\todoi{Page limit for all content is 16 pages! (currently \thepage)}
\fi

\end{document}